\documentclass{epl}

\usepackage{amssymb}

\title{Ground states versus low-temperature equilibria in\\
       random field Ising chains}
\shorttitle{Ground states vs. low-temperature equilibria in RF chains}

\author{G. Schr\"oder\inst{1,2} \and T. Knetter\inst{1,2} 
   \and M. J. Alava\inst{3} \and H. Rieger\inst{4}}
\shortauthor{G. Schr\"oder \etal}
\institute{
  \inst{1} Institut f\"ur Theoretische Physik, Universit\"at zu
           K\"oln, 50926 K\"oln, Germany\\
  \inst{2} NIC c/o Forschungszentrum J\"ulich, 52425 J\"ulich, Germany\\
  \inst{3} Helsinki University of Techn., Lab.\ of
           Physics, P.O.Box 1100, 02015 HUT, Finland\\
  \inst{4} Theoretische Physik, Universit\"at des Saarlandes,
           66041 Saarbr\"ucken, Germany
}
 {Lattice theory and statistics (Ising, Potts, etc.)}

\begin{document}

\maketitle

\begin{abstract}
  We discuss with the aid of random walk arguments and exact
  numerical computations the magnetization properties of
  one-dimensional random field chains.  The ground state structure is
  explained in terms of absorbing and non-absorbing random walk
  excursions.  At low temperatures, the magnetization profiles follow
  those of the ground states except at regions where a local random
  field fluctuation makes thermal excitations feasible. This follows
  also from the non-absorbing random walks, and implies
  that the magnetization length scale is a product of these two
  scales. It is not simply given by the Imry-Ma-like ground state
  domain size nor by the scale of the thermal excitations.
\end{abstract}

\newcommand{\be}{\begin{equation}}
\newcommand{\ee}{\end{equation}}

In statistical mechanics of random systems the search for universality
can be interpreted geometrically.  That is, if the introduction
of disorder into a system is relevant, the real-space properties of
the physical states can be understood in terms of scaling exponents.
These describe the fluctuations of a domain wall, or the behavior of a
spin-spin correlation function.  The central ingredient is that the
configurational energy is coupled to geometric fluctuations.
Consider a domain wall in a magnet. If the spatial fluctuations are
described by a roughness exponent $\zeta$, then there is an associated
exponent $\theta$ describing the free or ground state energy fluctuations. 
Assuming that
the 'zero temperature fixed point' scenario is true or that the
entropy is irrelevant at low enough temperatures, this is all what is
needed to describe the physics. The system evolves via Arrhenius-like
dynamics so that the cost of moving in the energy landscape is given
by the usual exponential factor $\exp(\Delta E \beta)$, where
$\beta=1/T$ and $T$ is the temperature, and $\Delta E \sim l^\zeta$
relates the cost to the scale length of the perturbation $l$.

Consider now a random magnet. It has a ground state (GS) which is described
exactly by the positions and arrangement of the domain walls. Examples
abound in particular in Ising systems, where non-trivial GSs
exist for spin glasses and random field systems \cite{review}.  In
this work we investigate with random walk arguments and exact
numerical computations how the aforementioned picture applies in the
case of one-dimensional random field chains.  We find that for
arbitrary field distributions \cite{Luck} the GS structure can be
understood via the random walk picture, which is compared to exact
numerical GS computations. At finite temperatures we resort to scaling
arguments based on this random walk picture, and again to an exact
numerical determination of the magnetization. This allows us to
make conclusions about the behavior in the same sample in both
cases. Our main finding is, in addition to recovering the GS from the
random walk picture, the emergence of two relevant length scales. These
arise from the zero-temperature length scale of the domains and the
typical size of 'easy' excitations at a given temperature. The latter
changes the correlation length of the magnetization, and thus leads to
the fact that in our case the low-temperature physics is characterized
not only by the zero-temperature scaling. The 1d RFIM has received
recently attention \cite{monthus} since it is simple enough that
decimation-type real space renormalization can be applied to domain
wall dynamics (each DW undergoes logarithmic Sinai diffusion
\cite{bouchaud}), which can be compared with our findings concerning
the asymptotic state of such processes. The zeroes of the
magnetization profile simply denote the equilibrium positions of
domain walls at $T>0$, and the extra physics consists of additional domain
walls added to the GS structure.  The chain is also the simplest
example of a random magnet with a competition of non-trivial GS and
thermal excitations (e.g. random bond Ising
magnets have a trivial GS).

\begin{figure}
\onefigure[width=9cm]{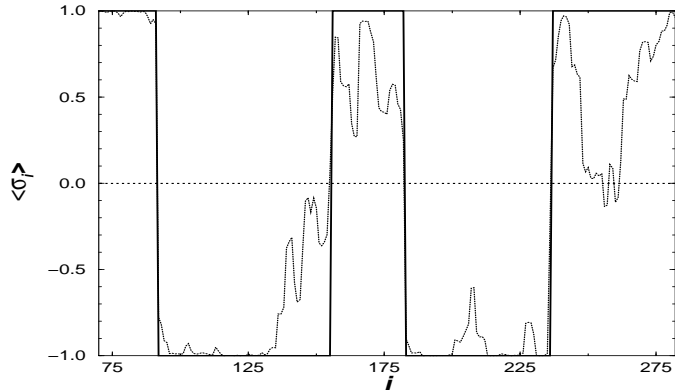}
\caption{Example for a ground state configuration (bold line) of a chain
  segment for a particular realization of random fields as compared to
  the equilibrium expectation values of the local magnetization
  $\langle\sigma_i\rangle$ at temperature $T=0.45J$. In the leftmost
  domain nearly no melting occurs, the region right of $i=125$ is a
  typical example for boundary melting and in the rightmost domain the
  bulk melting is so strong that even magnetization reversal occurs.}
\label{example}
\end{figure}

In the following we investigate the Hamiltonian
\begin{equation}
H = -\sum_{i=1}^N J \sigma_i \sigma_{i+1} + \sum_i h_i \sigma_i
\label{ham}
\end{equation}
where the $\sigma$'s are spins located at sites $i$ of the chain, and
$h_i$ are random fields picked from a suitable probability
distribution $P(h_i)$ with zero mean and variance $h_r$. For a binary
distribution $h_i=\pm h_{i,r}$ the model is equivalent to a spin glass
chain (with couplings $J_i=\pm J$) in a homogeneous external field $h_r$
\cite{sg,chenma}. Fig.\ \ref{example} shows  typical GS and
finite temperature ($T=0.4$) magnetization profiles obtained from the
numerical procedures described below. The GS domain size is
often thought to be given by the Imry-Ma argument \cite{imry},
which states that the domain field energy balances the cost from the domain 
walls on a scale $[l]_{\rm av} \sim 1/h_r^2$ in 1D and $[\ldots]_{\rm av}$
denotes the disorder average.  This reasoning omits the global
optimization behind the GS; later we discuss the exact way the
optimization becomes visible in. At finite but small temperatures the
magnetization changes due to two reasons. The
GS domain walls fluctuate, and thus the $m(x)$-profile is smoothed
out around the GS positions.  More interestingly, there are regions
inside domains where the magnetization can even undergo a local
reversal. Both of the cases arise from the local
random field configurations as we now demonstrate.

\begin{figure}
  \onefigure[width=7cm]{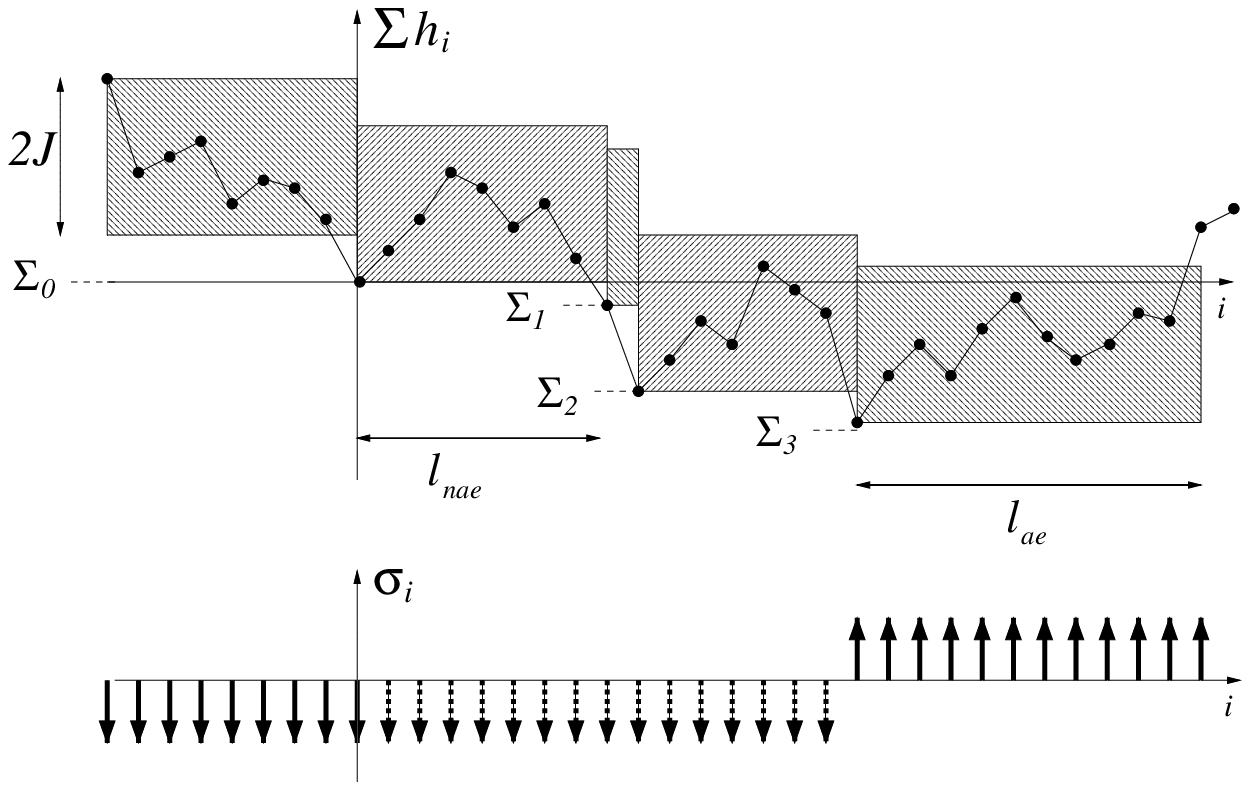}
  \caption{Terminology used for the description of the RW
           arguments. For further details see text.}
  \label{fig:GS.RW.illu_RW_arg}
\end{figure}

The starting point for the random walk argument is the fact that any
sequence $\mathcal{S}$ of lattice sites $i$ with $|\sum_{i\in
  \mathcal{S}} h_i|\ge 2J$ evidently leads to a GS spin structure with
$\sigma_i =+1 \quad \forall i \in \mathcal{S}$ if 
$\sum_{i\in \mathcal{S}} h_i\ge 2J$ (and 
$\sigma_i=-1\quad\forall i \in \mathcal{S}$ 
if $\sum_{\mathcal{i} \in S} h_i\le -2J$) independent of
the local fields $h_j$ at sites $j\not\in \mathcal{S}$.  The system
can thus be split up into such {\em absorbing excursions} and into
the remaining lattice sites, which make up so-called {\em non-absorbing 
excursions}.

Fig.~\ref{fig:GS.RW.illu_RW_arg} illustrates these concepts.
An {\em absorbing excursion}
is a sequence of spins starting at some lattice site $i$ and 
ending at the lattice site $j\ge i$, with the field-sum 
$|\sum_{i\in \mathcal{S}} h_i|$ for the first time becoming
greater or equal to $2J$:
\begin{equation}
  \label{eq:def_ae}
        |\sum_{l=i}^j h_i| \ge 2J \qquad \mbox{and} \qquad |\sum_{l=i}^k h_i| < 2J \quad \forall i < k < j.
\end{equation}
In Fig.~\ref{fig:GS.RW.illu_RW_arg} the left- and rightmost
sequences are absorbing excursions, of length $l_{ae}$. 
A sequence $\mathcal{S}'$ of spins from $i$ to $j\ge i$ 
is  a {\em non-absorbing excursion} if
\begin{equation}
  \label{eq:def_nae}
        \overline{\sigma}\sum_{l=i}^j h_i \le 0 \qquad \mbox{and} \qquad 0 <\overline{\sigma}\sum_{l=i}^k h_i < 2J \quad \forall i < k < j 
\end{equation}
where $\overline{\sigma}=\pm 1$ is the orientation of the spins within the 
preceding absorbing excursion. The length of a non-absorbing excursion is 
$l_{nae}$. A simple 'step down' (like from $\Sigma_1$ to $\Sigma_2$) 
is included in this definition.

The GS now follows as a sequence of absorbing and non-absorbing
excursions. It, and the Zeeman energy and mean domain-length can be
determined with the three rules: (1) determine an absorbing excursion
$\mathcal{S}_0$ for a given field configuration. If it starts at site
$i_0$, ends at $j_0$, and $\overline{\sigma}$ is the sign of its
field-sum, then $\sigma_k=\overline{\sigma}$ for all $k\in
\mathcal{S}_0$. (2) start from $j_0+1$ and find all $n_{nae}$ of
non-absorbing excursions until the next absorbing excursion
$\mathcal{S}_1$ (from $i_1$ to $j_1$) is found, whose field-sum is by
definition opposite in sign to the preceding one. The sites $k$
belonging to the non-absorbing excursions have the same orientation
$\sigma_k=\overline{\sigma}$ as those within $\mathcal{S}_0$. The
orientation of the spins at sites $l$ within $\mathcal{S}_1$ is
opposite to the latter one, $\sigma_l=-\overline{\sigma}$.  (3)
starting again at $j_1+1$ the search (2) for the next absorbing
excursion then leads to the overall GS.  These steps actually define a
fast algorithm for finding the GS, though for historical reasons we
have used the mapping to the max-flow/min-cut problem \cite{review}.

\begin{figure}
\twofigures[width=0.47\textwidth]{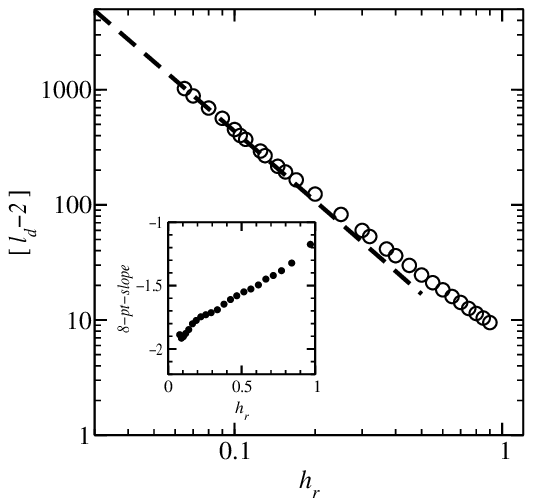}{P_Ld_gau.eps}
\caption{Average domain length as a function of $h_r$. The dotted line is
  a fit to eq.~(\ref{eq:mean_ld_concrete}) with $a=-0.74$, $b=0.25$
  and $c=1.4$. The inset shows the 8-point slope of the presented data
  again yielding an exponent $2$ in the limit of small field
  amplitudes.}
\label{fig:ld_vs_h}
\caption{Probability distribution of the domain lengths $l_d$. 
  Apart from a non-exponential tail which might be due to finite size effects
  the decay is exponential with decay rate $\nu$.  {\bf Inset:} The decay rate
  $\nu$ as a function of $h_r$. For $h_r\ll 1$ the data are compatible with
  $\nu\sim h_r^2$ (bold line).}
\label{fig.P(ld)}
\end{figure}

The mean domain length $[l_d]_{\rm av}$ is given by 
\begin{equation}
  \label{eq:mean_ld_abstract}
        [ l_d ]_{\rm av} (h_r)=[ n_{nae} ]_{\rm av} 
        [ l_{nae}]_{\rm av} + [ l_{ae} ]_{\rm av}.
\end{equation}
Thus the domain length consists of two distinct contributions and we
need to estimate the $h_r$-dependence of $[n_{nae}]_{\rm av}$,
$[l_{nae}]_{\rm av}$ and $[l_{ae}]_{\rm av}$. The fieldsum over the
local fields of a non-absorbing excursion is a RW with absorbing
boundaries at $\sum_{nae}h_i=0$ and $\sum_{nae}h_i=2J$ and random step
size with zero mean and variance $h_r$. Rescaling the step size
$h_i'=h_i/h_r\rightarrow 1$ this becomes a 1d RW starting from $x=1$
at $t=0$ with random step lengths with mean zero and variance one (for
a binary distribution $h_i=\pm h_r$ it yields a conventional lattice
random walk with $h_i'=\pm1$) and absorbing boundaries at $x=0$ and
$x=L=2J/h_r$. The probability $P_0(t,L)$ to be absorbed at $x=0$
within the time interval $[t,t+dt]$ without having been absorbed at
$x=L$ reads \cite{ir} $P_0(t,L)\propto t^{-3/2}$ for $1\lesssim t
\lesssim L^2$ and decays exponentially for $t\gtrsim L^2$.
Integration over $P_0(t,L)$ leads to $[l_{nae}]_{\rm av} \sim L\propto
h_r^{-1}$.  The mean number $[n_{nae}]_{\rm av}$ of consecutive
non-absorbing excursions follows due to the fact that the probability
for an excursion to be absorbing grows as $p_{ae}\sim 1/L\propto h_r$
\cite{ri}.  Thus $P(n_{nae})\sim (1-p_{ae})^{n_{nae}}$ decays
exponentially.  As a consequence $n_{nae}\sim 1/{\ln(1-b\,h_r)}$ and
the mean length of an absorbing excursion grows like $h_r^{-2}$.
Finally eq. (\ref{eq:mean_ld_abstract}) reads
\begin{equation}
  \label{eq:mean_ld_concrete}
        [ l_d ]_{\rm av} (h_r)\sim \frac{a}{h_r\,\ln(1-b\, h_r)}
        +\frac{c}{{h_r}^2}\quad \rightarrow \quad
        \frac{e}{{h_r}^2}\qquad 
        \mbox{for} \quad h_r\rightarrow 0
\end{equation}
where one expects $b< 1$, $a<0$ and $-a\approx c$. Note that
$a/h_r\,\ln(1-b\, h_r)\sim h_r^{-2}$ for $h_r\rightarrow 0$ and for
$h_r<0.5$ no significant difference between $a/[h_r\ln(1-b\, h_r)]$
and $h_r^{-2}$ can be observed. The asymptotic limit
follows the Imry-Ma scaling, though
the physics is more complicated. 

This result is confirmed by computations of exact ground states.
The data was obtained for a Gaussian random field distribution with
zero mean and variance $h_r$ and averaged over $10^5$ disorder
configurations.  The system size is large enough ($L=5000$) such that
$[l_d]_{\rm av}\ll L$ even for the smallest field strength $h_r$.
Fig.~\ref{fig:ld_vs_h} shows our numerical result for the average
length $[l_d-2]_{\rm av}$ of the GS domains as a function of the field
amplitude $h_r$.  In the limit $h_r\rightarrow \infty$ $[l_d]_{\rm av}
\rightarrow 2$ since all the spins align with their local fields.  In
the limit $h_r\ll J$ the data fit well to the predicted form
eq.~(\ref{eq:mean_ld_concrete}), scaling as $h_r^{-2}$ for
$h_r\rightarrow 0$.  Moreover, as can be seen in Fig.\ \ref{fig.P(ld)}
the probability distribution of the domain sizes decays exponentially,
with a decay rate $\nu$ that scales inversely proportional to
$[l_d]_{\rm av}$, i.e.\ $\nu(h_r)\propto h_r^2$.

\begin{figure}
\twofigures[width=0.47\textwidth]{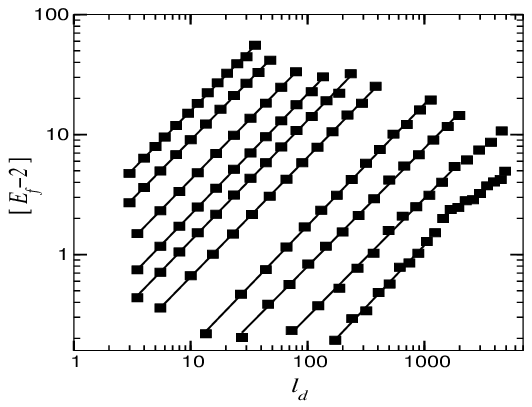}{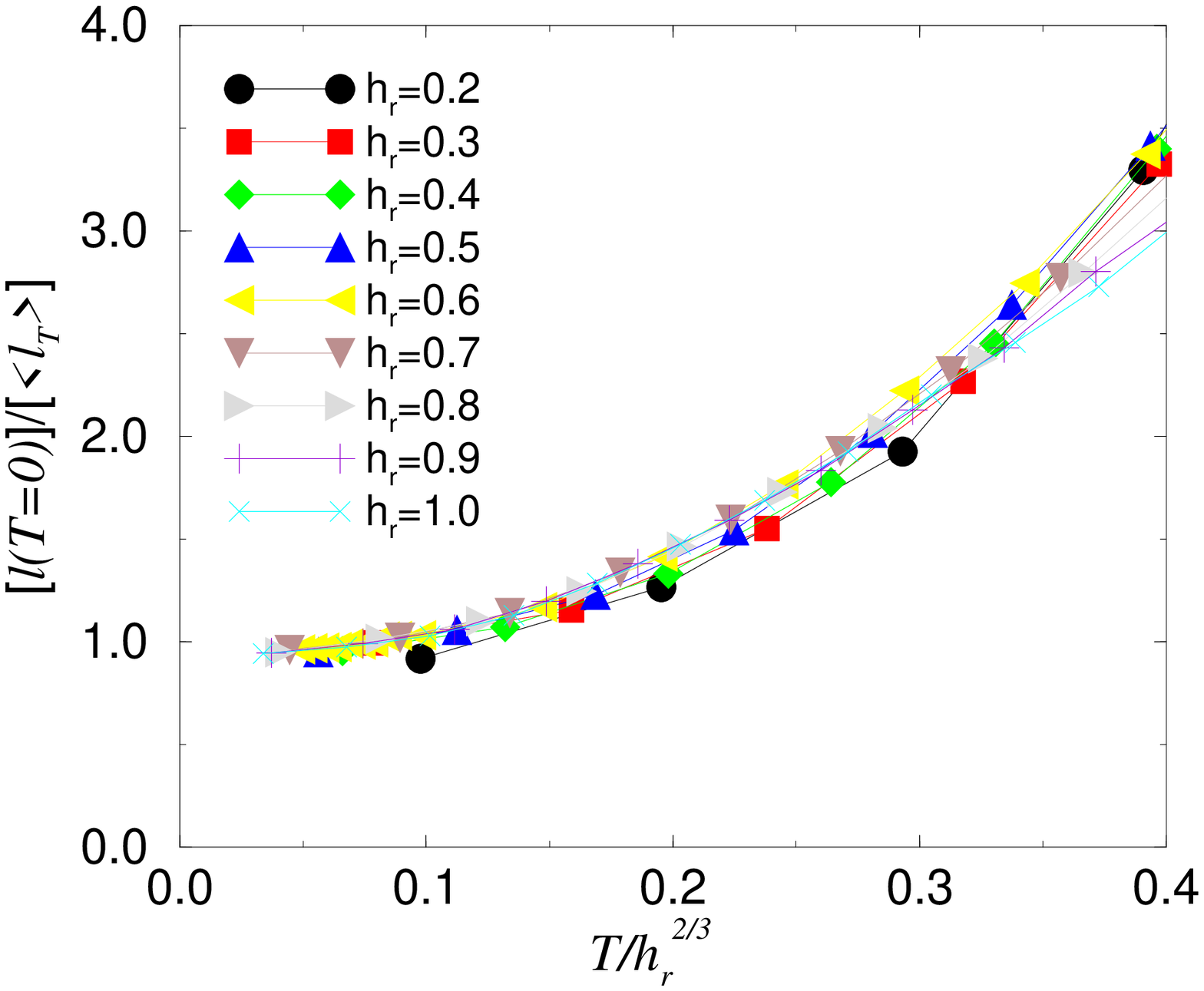}
\caption{
  Mean Zeeman energy $H_Z-2J$ corresponding to a particular domain
  length $l_d$ in a log-log plot. $h_r=3.0,2.0,1.2,
  1.0,0.8,0.6,0.4,0.2,0.13,0.08,0.05$ from top to bottom. The slopes
  of the straight lines are all within the interval $1.00\pm 0.05$.
  The data are averaged over $10^5$ disorder configurations. The
  straight lines represent least square fits.}
\label{fig:H_Z_original-data}
\caption{Scaling-plot for the average domain-lengths at
  finite temperatures: $[l(T=0)]_{\rm av}/[\langle l_T\rangle]_{\rm
    T}$ versus $T/\xi(h)$ for different values of the field strength
  $h_r$, where the length scale $\xi(h)\propto h_r^{2/3}$.}
\label{fig:domlength}
\end{figure}

The field energy of a domain can be computed as a function of $h_r$
and $l_d$ by noting that both a single absorbing excursion and all of
the non-absorbing excursions contribute. The former contributes
a constant ($2J$), depending neither on $h_r$ nor on $l_d$. Each
non-absorbing excursion adds an amount of $\mathcal{O}(h_r)$ so that
the sum self-averages. The contribution of a single non-absorbing
excursion equals $\Sigma_i-\Sigma_{i-1} \sim h_r$, i.e.  the step
width of the RW. Thus the field energy results from the number of
non-absorbing excursions in a domain, $n_{nae}$, plus $2J$. From
(\ref{eq:mean_ld_concrete}) we learn that in the limit $h_r\to\infty$
the contribution of the absorbing and non-absorbing walks to
$[l_d]_{\rm av}$ scale similarly such that we expect that for a fixed
domain size $[n_{nae}(l_d)]_{\rm av}\propto l_d/[l_{nae}]_{\rm
  av} \propto l_d h_r$.  Thus
\begin{equation} 
[E_f(l_d)]_{\rm av} 
= 2J + [n_{nae}(l_d)]_{\rm av}\,h_r = 2J + d\,h_r^2\,[l_d]_{\rm av}
\end{equation}
The numerics confirms this result: Fig.
\ref{fig:H_Z_original-data} shows the data for the mean Zeeman energy
$[E_f(l_d)]_{\rm av}$ of domains of length $l_d$. From the slopes of the
straight lines we learn that $[E_f(l_d)]_{\rm av}$ is linear in the domain
length and from the offsets that it grows like $h_r^2$, independent of
the field distribution $P(h_i)$. Note that from a {\it naive} random walk
picture one would expect $[E_f(l_d)]_{\rm av}\propto l_d^{1/2}h_r$, 
which is incorrect.

We now turn our attention to equilibrium configurations, i.e.\ the
local magnetization $m(x)$ and the domain structure at $T>0$.  Using
numerical transfer matrix methods [10,11] to compute the partition
function $Z_N$ we can compute the exact expectation value $\langle
\sigma_r \rangle$ for each spin $\sigma_r$ by calculating the product
of the $N$ $2\times 2$ transfer matrices.  Since some of the random
matrix elements can be very small, floating point accuracy gives a
lower limit of $T=0.05$.

First we address the scale-lengths of the equilibrium magnetization by
computing the average length $[l_m]_{\rm av}$ that separates two zeroes
of the magnetization $m(x)$. Figure \ref{fig:domlength} demonstrates how this
length-scale changes with temperature, if we first scale away the
$T=0$-dependence on the field. A further collapse with the right combination
of $h_r$ and $T$ makes it possible to observe an universal scaling function
for $[l_m]_{\rm av}$
\be
[l_m]_{\rm av} = [l_d]_{\rm av} f(T/h_r^{2/3}),
\ee
where the scaling function $f \rightarrow 1$ with $T\rightarrow 0$.
The dependence of $[l_m]_{\rm av}$ on the combination of temperature
and field strength does not follow an Imry-Ma-like scaling but is a
consequence of entropic effects. The length $[l_m]_{\rm av}$ at
a finite temperature is determined by both a zero-temperature scale
($[l_m]_{\rm av}$) and thermal fluctuations. The following argument
can explain the scaling variable $h_r^{2/3}/T$, analogously to spin
glass chains in an external field \cite{chenma}. Once again
consider the non-absorbing random walks which the domains consist of.
Some of these inside a typical domain are such that the random walk
sum of fields over the excursion is close to $2J$.  These
almost-absorbing walks are the sequences (of spins) most likely to be
flipped at finite temperatures.  The cost of flipping such a part of a
domain is proportional to $J$, which if measured in terms of $h_r$ can
be written with the help of the length-scale $l$ of the non-absorbing
excursions, $h_r \sim 1/l$.  This is {\it almost} equal to the
Zeeman-energy optimized over the excursion, which scales as $E_f \sim
h_r l^{1/2}$. Equating the cost with the gain  and solving for the
energy scale ($E_f$) as a function of field gives rise to the
Arrhenius factor $E_f/T \sim h_r^{2/3}/T$.

\begin{figure}
\twoimages[width=0.47\textwidth]{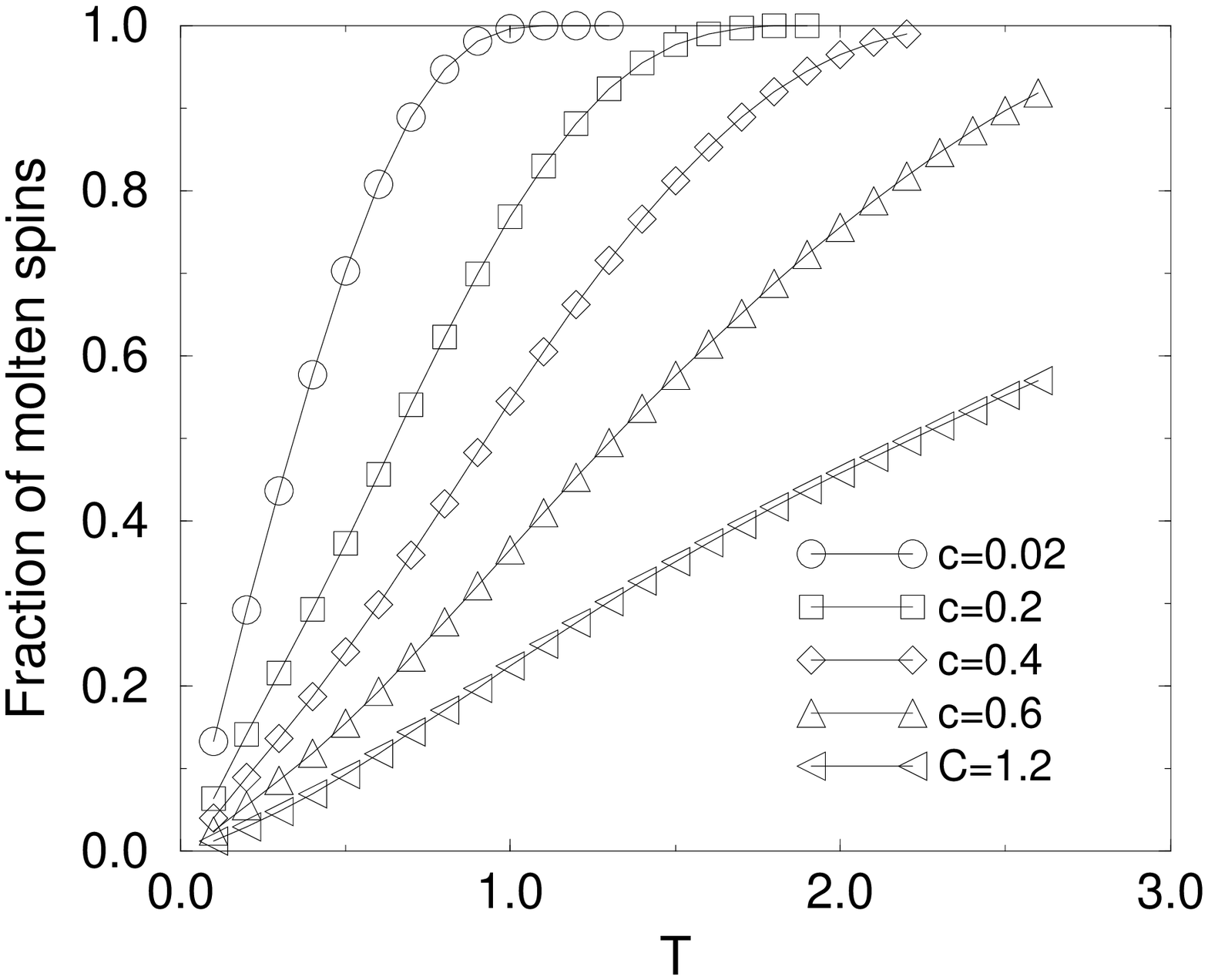}{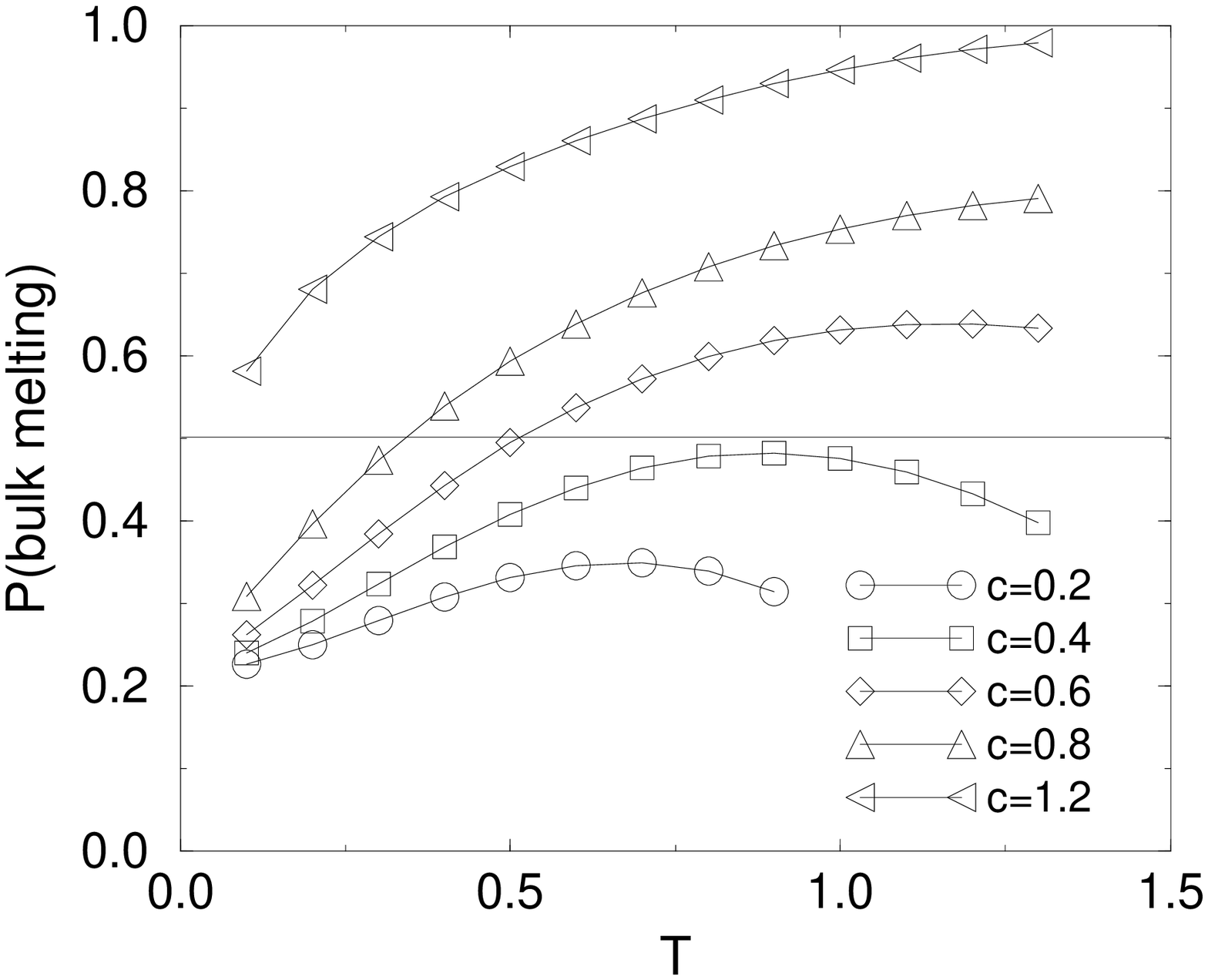}
\caption{{\bf Left:} Fraction of melted spins as a function of temperature
  for different values of $c$.\newline
  {\bf Right:} Portion of those spins is
  displayed which reside inside bulk segments.  $L=400$, $h_r=1.0$.}
\label{fig:propdist_T}
\label{fig:melted_spins}
\end{figure}

As Fig. 1 demonstrates, the magnetization profile at $T>0$ differs
from the GS due to domain wall fluctuations {\it and} internal cluster
reversals. To study this quantitatively we introduce a parameter
$c\in(0,2)$ and define a reversal to be a sequence of spins for which
$ |\langle \sigma_i \rangle (T)-\sigma_i(T=0)| > c$ holds; moreover
the definition can be applied to both processes separately revealing
interesting details.  Since bulk reversal is always coupled with the
breaking of two extra bonds one expects that domain wall fluctuations
dominate. However, the former contributes a considerable portion to
the total melting even at low temperatures
(Fig.~\ref{fig:propdist_T}). The relative portion of bulk reversals at
first grows with temperature for all values of $c$ since the gain in
entropy allows for more broken bonds. Moving the threshold $c$ away
from $1$ and $-1$ respectively, a greater number of bulk segments are
identified. Eventually for very large $c$ even more bulk than boundary
reversals are observed. The characteristic reversal rates are
different for the two processes, and related via the empirical formula
\begin{equation}
  \label{eq:factor}
  (\frac{\Delta m}{\Delta T})_{bulk} 
  = \alpha (\frac{\Delta m}{\Delta T})_{bound}\;,
\end{equation}
with $\alpha\!\approx\!1.63$. Thus the change in magnetization with
increasing $T$ is stronger inside the GS domains than at their
boundaries. These results are independent of the field strength.

In conclusion, we have studied the magnetization properties of
one-dimensional RFIM chains. These can be explained using of random
walk arguments. While the GS structure is found to be a sequence of
absorbing and non-absorbing excursions, the finite-temperature
magnetization is complicated by thermal excitations. These are
explained with the help of almost-non-absorbing walks. The results
illustrate how a global optimization problem influences physics at
$T>0$ in systems where the geometric arrangement of domain walls is
crucial. Yet, extending the results to higher dimensions seems
insolvable.

\acknowledgments
This work has been supported by the Academy
of Finland and the German Academic Exchange Service (DAAD)
within a common exchange project, and separately, by
the A. of F.'s Center of Excellence Program.

\end{document}